%% file: htocsbar.tex
\documentclass[a4paper]{PoS}
\usepackage{color}
\RequirePackage{xspace}
\usepackage{relsize}
\def\CP     {\ensuremath{C\!P}\xspace}
\def\Bbar   {\kern 0.18em\overline{\kern -0.18em B}{}\xspace}
\def\Dbar   {\kern 0.2em\overline{\kern -0.2em D}{}\xspace}
\def\Kbar   {\kern 0.2em\overline{\kern -0.2em K}{}\xspace}
 
\newcommand {\gev}{\ensuremath{{\mathrm{\,Ge\kern -0.1em V}}}\xspace}
\newcommand {\tev}{\ensuremath{{\mathrm{\,Te\kern -0.1em V}}}\xspace}
\newcommand {\gevcc}{\ensuremath{{\mathrm{\,Ge\kern -0.1em V\!/}c^2}}\xspace}
\def\pb     {\ensuremath{\mbox{\,pb}}\xspace}
\def\invfb  {\ensuremath{\mbox{\,fb}^{-1}}\xspace}

\title{Search for a light charged Higgs boson decaying into {\boldmath $c\bar{s}$} at CMS}

\ShortTitle{Search for $H^+\to c\bar{s}$ at CMS}

\author{\speaker{Gouranga Kole}\thanks{On behalf of the CMS Collaboration}\\
        Tata Institute of Fundamental Research\\
        Homi Bhabha Road, Colaba\\ 
        Mumbai 400005, India\\
        E-mail: \email{gouranga@tifr.res.in}}

\abstract{We present results on the search for a light charged Higgs boson that can be
produced in the decay of a top quark and later decays into a charm and an antistrange
quark. The analysis is performed using $19.7\invfb$ pp collison data recorded with the
CMS detector at LHC.}

\FullConference{Prospects for Charged Higgs Discovery at Colliders - CHARGED 2014,\\
		16-18 September 2014\\
		Uppsala University, Sweden}

\begin{document}

\section{Introduction}
A Higgs boson has recently been discovered by ATLAS~\cite{Aad:2012tfa} and
CMS~\cite{Chatrchyan:2012ufa} with a mass close to $125\gev$ and properties,
within uncertainties of the available data, consistent with those expected from
the standard model (SM). Although this would complete the SM, still the latter
cannot be the full story having many missing links such as dark matter, baryon
asymmetry and gravity. Several extensions to the SM have been proposed to address
these inconsistent features. The minimal supersymmetric standard model (MSSM) is
one such model that contains two Higgs doublets, resulting in five physical Higgs
states: a light and heavy $\CP$-even $h$ and $H$, a $\CP$-odd $A$, and two charged
Higgs bosons $H^{\pm}$. At tree level, the MSSM Higgs sector can be expressed
in terms of two parameters, which are usually chosen to be the mass of the
$\CP$-odd Higgs boson ($m_A$) and the ratio of the vacuum expectation values
of the two Higgs doublets ($\tan\beta$).

The lower limit on the charged Higgs boson mass is $78.6\gev$, as determined by
LEP experiments~\cite{Achard:2003gt,Heister:2002ev}. If the mass of the charged
Higgs boson is smaller than the mass difference between the top and the bottom
quarks, the top can decay via $t\rightarrow H^{+}b$.
For $\tan\beta <1$, the charged Higgs boson preferentially decays to a charm
and an antistrange quark ($c\bar{s}$).
In the two Higgs doublet model of type I and Y the branching fraction
${\cal B}(H^+\to c\bar{s})$ is larger than $10\%$ for any value of $\tan\beta$,
while in type II and X it can reach up to $100\%$ for $\tan\beta <1$~\cite{PhysRevD.80.015017}.
In this study, we assume ${\cal B}(H^+\to c\bar{s})$ to be $100\%$. Recently ATLAS has 
set an upper limit on ${\cal B}(t\to H^{+}b)$ between $5\%$ and $1\%$ for charged Higgs 
masses in the range $90$-$150$ GeV~\cite{atlashtocs}.

\begin{center}
\begin{figure}[htp]
\includegraphics[width=0.55\textwidth]{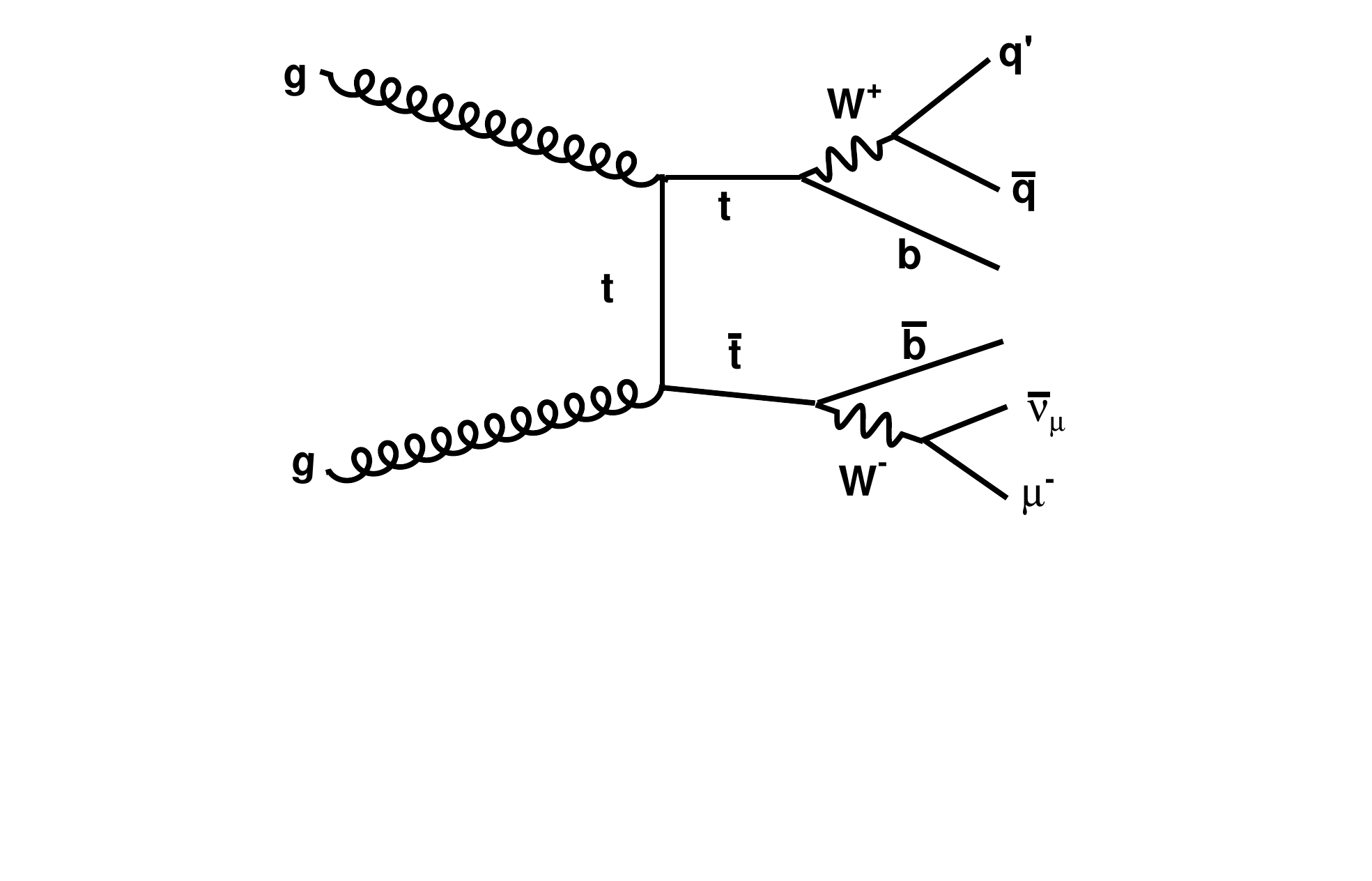} 
\includegraphics[width=0.55\textwidth]{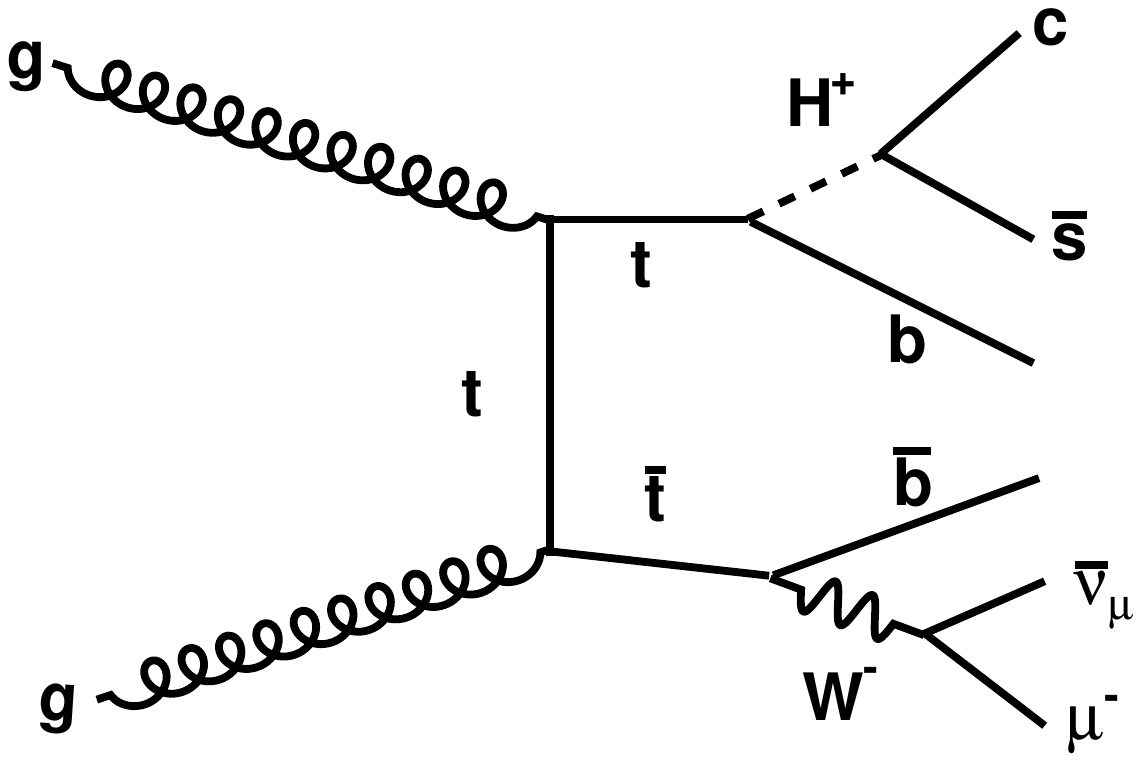}
\caption{Leading-order Feynman diagrams for (left) the SM $t\bar{t}$ production
at LHC in the muon final state, and (right) the same in presence of the charged
Higgs boson~\cite{CMS-HIG-13-035}.}
\label{fig:feynHiggs}
\end{figure}
\end{center}

The presence of the $t\to H^+b$, $H^+\to c\bar{s}$ decay channel alters the
event yield for $t\bar{t}$ pairs having hadronic jets in the final state,
compared to the SM prediction. The search for a charged Higgs boson is thus
sensitive to the decays of the top pairs $t\bar{t}\to H^{\pm}bW^{\mp}\bar{b}$
 and $t\bar{t}\to H^{\pm}bH^{\mp}\bar{b}$, where the charged Higgs boson decays
into a charm and an antistrange quark. We perform a model independent search~\cite{CMS-HIG-13-035}
for the charged Higgs boson in the $t\bar{t}\to H^{\pm}bW^{\mp}\bar{b}\to\mu+
E^{\rm miss}_T+{\rm jets}$ final state, where the $W$ boson decays to a muon and
a neutrino (leading to missing transverse energy $E^{\rm miss}_T$) and the $H^+$
decays to $c\bar{s}$. The contribution of the process $t\bar{t}\to H^{\pm}bH^{\mp}
\bar{b}$ is expected to be negligible in the above final state.
Figure~\ref{fig:feynHiggs} shows the dominant Feynman diagrams for this final
state both in the SM $t\bar{t}$ process as well as the same in presence of the
$H^+$ boson.

\section{CMS Detector and Object Reconstruction}
The distinguishing features of the CMS detector~\cite{detectorCMS} are a 6\,m
long solenoidal magnet that produces 3.8 T magnetic field, a fully silicon-based
tracking device, a $PbWO_4$ crystal electromagnetic calorimeter, a brass-scintilltor
sandwich hadron calorimeter, and an excellent muon system.
All physics objects used in the analysis are reconstructed with the particle flow
(PF) algorithm, essentially combining information from the aforementioned
subdectectors. Muons are reconstructed by matching the tracks in the silicon
tracker with the hits in the muon system. Jets are reconstructed based on the
anti-$k_T$ algorithm with a cone radius parameter $R=0.5$. The $E^{\rm miss}_T$ is
defined as the negative vector sum of the transverse momenta ($p_T$) of all PF candidates. 
To identify jets originating from a $b$ quark, we apply the $b$-jet identification
criteria that involve the use of secondary vertices together with track-based
lifetime information.

\section{Backgrounds}
Different backgrounds such as $t\bar{t}$, $W$+jets and $Z$+jets are generated with
{\textsc Madgraph} 5~\cite{Alwall:2011uj} interfaced with {\textsc Pythia} 6.4~\cite{Sjostrand:2006za}.
The UE tuning Z2*~\cite{z2tune} and CTEQ6M~\cite{cteq} PDFs are used. The $t\bar{t}$
events are estimated from the next-to-next-to-leading-order (NNLO) SM prediction with
a production cross section of $245.8\pm 6.0\pb$. The single top processes are calculated
using {\textsc Powheg}~\cite{powheg}.
The $W$+jets background is calculated at NNLO with FEWZ3.1~\cite{fewz}, while $Z$+jets
and single top events are also normalized to NNLO cross-section calculations.
The signal $t\bar{t}\to H^{\pm}bW^{\mp}\bar{b}$ sample is generated with {\textsc Pythia} 6.4
and normalized using the same production cross section as the SM $t\bar{t}$. The cross
sections of diboson backgrounds ($WW$, $WZ$ and $ZZ$) are computed with MCFM~\cite{mcfm}.

\section{Event Selection and Analysis}
We select events with a well-identified muon having $p_T>25\gev$ and pseudorapidity
$|\eta|<2.1$. The muon is required to be isolated from the surrounding hadronic activity
by imposing a PF-based relative isolation criterion, $I_{rel}<0.12$~\cite{CMS-HIG-13-035}. Any
event that contains an additional muon or electron with $p_T>10\gev$ and $|\eta|<2.5$ passing
a loose isolation requirement, $I_{rel}<0.3$, is rejected. For a consistent data-MC matching,
the trigger, muon ID and isolation scale factors are applied to MC events as a function of
muon $p_T$ and $\eta$. These scale factors are derived using a tag and probe technique based
on $Z\to\mu\mu$ events. Events are required to have at least four jets with $p_T>30\gev$ and
$|\eta|<2.5$, where two of them arise from top quarks and the other two from $W$/$H^{\pm}$
boson. Owing to the missing neutrino in the final state, we require the event to have
$E^{\rm miss}_{T}>20\gev$. This criterion also helps
in suppressing the QCD multijet and $Z$+jets backgrounds, where there is no real source of
$E^{\rm miss}_{T}$. A kinematic fit is used to reconstruct $t\bar{t}$ events from the final
states resulting in an improved mass resolution for the hadronically decaying boson. This fit
constrains the event to the hypothesis for a production of two top quarks, each one decaying
to a $W$ boson and a $b$ quark. As indicated above, one of the $W$ bosons decays into a
muon-neutrino pair, while the other boson ($W$ in SM $t\bar{t}$ or $H^{+}$ for the signal)
decays into a quark-antiquark pair. As we are interested in the reconstruction of the $W$/$H^{+}$
boson mass, we constrain the reconstructed mass of the two top quarks to $172.5\gev$ in the fit.
Only jets passing the $b$-tagging requirement are considered as candidates for the $b$ quarks
in the $t\bar{t}$ hypothesis, while all other jets are considered as candidates for the light
quarks in the boson's hadronic decay. For each event, the assignment that gives the best fit
probability is finally retained. As the top $p_T$ spectrum is found to be softer in simuations
compared to the data, we reweight the $t\bar{t}$ MC events according to the scale factors
derived based on Ref.~\cite{toppt}.

\begin{figure}[htbp]
\begin{center}
\includegraphics[width=0.42\textwidth]{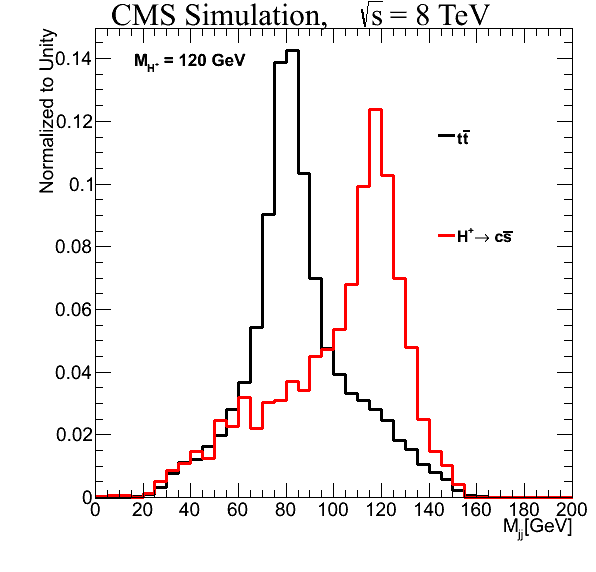}
\includegraphics[width=0.42\textwidth]{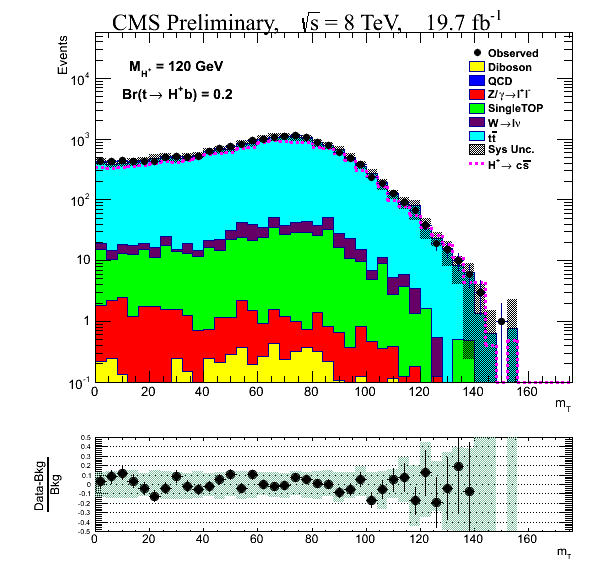}
\caption{(Left) Invariant mass distributions of the dijet system decaying to $c\bar{s}$ obtained
from a kinematic fit for the SM $t\bar{t}$ events and the same in the presence of $H^+$ ($m_H=120\gev$)
after all selections with MC simulated samples. (Right) the $m_T$ distribution after the kinematic
fit and all other selections~\cite{CMS-HIG-13-035}.}
\label{fig:dijet_mass_comp_wmt}
\end{center}
\end{figure}

Figure~\ref{fig:dijet_mass_comp_wmt} (left) shows the $W$ and $H^{+}$ boson mass distributions
obtained from the kinematic fit after full event selection with MC simulated samples. As evident
from the plot, the kinematic fit significantly improves the dijet mass resolution, which is
crucial in separating the $H^+$ from the $W$ peak. Figure~\ref{fig:dijet_mass_comp_wmt} (right)
shows the transverse mass ($m_T$) distribution of the system comprising the muon and $E^{\rm miss}_T$,
which shows a good agreement between data and expected SM background from MC simulations.

\section{Systematic Uncertainties}
\label{s:systematics}
The uncertainty in the jet energy scale (JES) is the leading source of systematic uncertainty
in this analysis, since we are using the dijet mass as the final discriminator. It is
estimated as a function of the jet $p_T$ and $\eta$ according to Ref.~\cite{Chatrchyan:2011ds},
and is then propagated to $E^{\rm miss}_T$. The uncertainty in JES changes both the event yield
and shape of the dijet ($W$ or $H^+$) mass distribution. To estimate the uncertainty in the dijet
mass distribution, the jet momenta are scaled according to the JES uncertainty by $\pm 1\sigma$.
The difference in the dijet shape used as a shape uncertainty in the limit calculation. The
uncertainty in $b$-tagging efficiency and light-jet misidentification probability is another
major source of uncertainty, as our selection requires two $b$-tagged jets. The theoretical
uncertainties on the cross sections of various processes are also considered. The uncertainty
in the production cross section of the $t\bar{t}$ process, which is common to both SM $t\bar{t}$
and signal channel, is an important source of uncertainty. The uncertainty due to the variation
of renormalization and factorization scales used in $t\bar{t}$ simulations is studied by
simultaneously changing the nominal scale values by factors of $0.5$ and $2.0$. An additional
shape nuisance is also taken into account as the uncertainty due to matching thresholds used
for interfacing the matrix elements generated with {\textsc Madgraph} and {\textsc Pythia} parton
showering. The thresholds are changed from the default value of $20\gev$ down (up) to $10$ ($40$)
$\gev$. Other bin-by-bin uncertainties are also applied to all backgrounds. Further, a nuisance
parameter corresponding to the uncertainty in the top $p_T$ reweighting is considered. Finally,
the uncertainty in the integrated luminosity is estimated to be $2.6\%$.

\section{Results and Summary}
The event yields after all selections are listed in Table~\ref{tab:mu_final_cutflowtable_incl}
along with their statistical and systematic uncertainties. The expected number of signal events
from the $t\bar{t}\to H^{\pm}bW^{\mp}\bar{b}$ ($HW$) process for ${\cal B}(t\to H^{+}b)=20\%$ is
also presented in the table.
\begin{table}[htbp]
\small\addtolength{\tabcolsep}{-5pt}
\begin{center}
\caption{Expected signal and background events are provided with their statistical and systematic
uncertainties. The number of events observed in the $19.7\invfb$ of data is also presented.}
\label{tab:mu_final_cutflowtable_incl}
\input{tabs/mjj_kfit_Id_incl_cutflow.tex}
\end{center}
\end{table}
Assuming that any excess or deficit of events in data, when compared with the expected background
contribution, is due to the potential signal $t\to H^+b$, $H^+\to c\bar{s}$ decay, the difference
between the observed number of data events and the predicted background contribution ($\Delta N$)
can be given as a function of $x={\cal B}(t\to H^+b)$ via:
\begin{equation}
\Delta N=N^{obs}_{t\bar{t}}-N^{SM}_{t\bar{t}}= 2x(1-x)N^{HW}_{t\bar{t}} + [(1-x)^2-1]N^{SM}_{t\bar{t}}
\label{eqn:pas_deltaN}
\end{equation}
Here, $N^{HW}_{t\bar{t}}$ is estimated from MC simulations forcing the first top quark to decay to
$H^{\pm}b$ and the second one to $W^{\mp}b$, and $N^{SM}_{t\bar{t}}$ is also calculated based on
simulations, as given by the SM $t\bar{t}$ entry in Table~\ref{tab:mu_final_cutflowtable_incl}.
Note that Eq.~(\ref{eqn:pas_deltaN}) is applicable to any new physics model as there is no explicit
dependence on any MSSM parameter. Therefore, our obtained limit in absence of a significant excess
or deficit will be model independent in nature.

\begin{figure}[htbp]
\begin{center}
\includegraphics[width=0.43\textwidth]{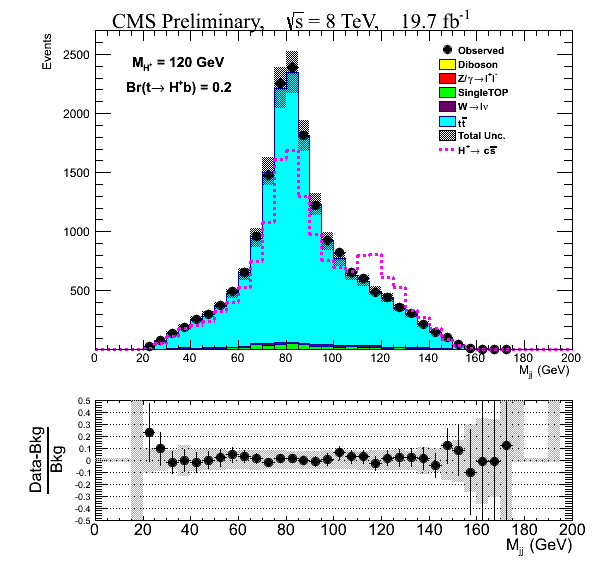}
\includegraphics[width=0.47\textwidth]{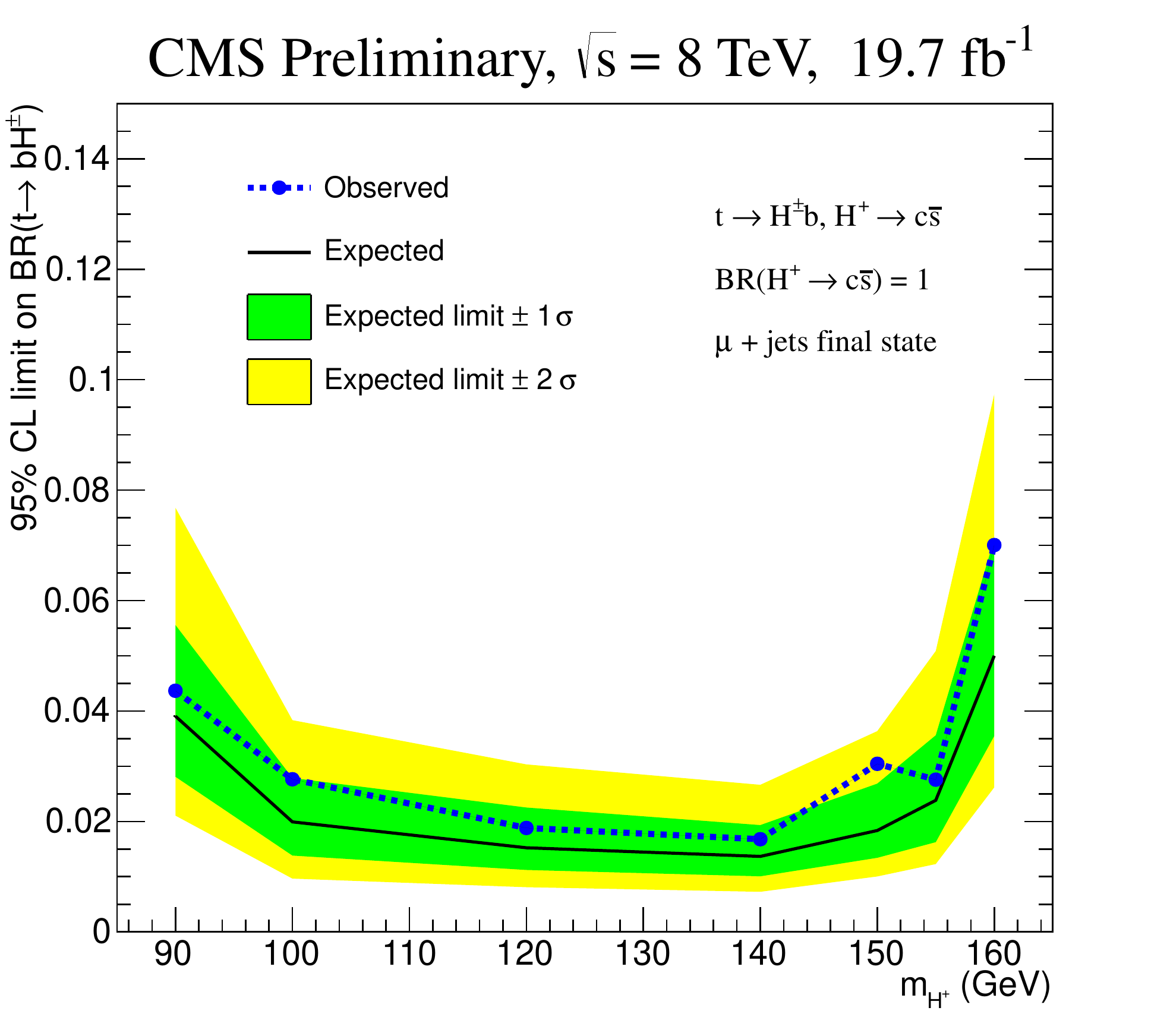}
\caption{(Left) Dijet mass distributions of the hadronically decaying boson after the maximum likelihood fit. (Right) Exclusion limit on ${\cal B}(t\to H^+b)$ as a function of $M_{H^+}$ assuming ${\cal B}(H^+\to c~\bar{s})=100\%$~\cite{CMS-HIG-13-035}.}
\label{fig:limit_incl_and_dijet}
\end{center}
\end{figure}

The LHC-wide CLs method~\cite{CLs,Junk:1999kv} is used to obtain an upper limit on $x={\cal B}(t\to H^{+}b)$
at the $95\%$ confidence level (CL) using Eq.~(\ref{eqn:pas_deltaN}). The dijet mass distribution shown in
Fig.~\ref{fig:limit_incl_and_dijet} is used in a binned maximum-likelihood fit to extract a possible signal.
The upper limit on $\cal{B}$($t\to H^{+}b$) as a function of $m_{H^{+}}$ is shown in Fig.~\ref{fig:limit_incl_and_dijet}.
The observed limit agrees well with the expected limit within one standard deviation. 

A search has been performed for a light charged Higgs boson produced in a top quark decay, subsequently 
decaying into $c\bar{s}$. The total integrated luminosity of $19.7\invfb$ recorded by CMS at $\sqrt{s}=8\tev$
are used in the search. 
In absence of any signal on the dijet invariant mass distribution of the $H^+\to c\bar{s}$ candidate events,
we set a model-independent $95\%$ CL upper limit on the branching fraction ${\cal B}(t\to H^+b)$ assuming
${\cal B}(H^+\to c\bar{s})=100\%$. The obtained limits are in the range of $2-7\%$ for a charged Higgs mass
between $90$ and $160\gev$.

\acknowledgments
I thank the organizers and CMS collaboration for giving me an opportunity to deliver a talk at this conference.

\end{document}

%% file: tabs/mjj_kfit_Id_incl_cutflow.tex
\begin{tabular}{ | c| c| }
\multicolumn{2}{c}{ } \\
\hline 
\multicolumn{1}{|c|}{Source} & \multicolumn{1}{|c|}{N$_{\rm events}$ $\pm$ uncertainty } \\
\hline 
\hline 
{\it HW}, $M_{H}=120$~GeV, ${\cal{B}}(t \to bH^{+}) = 20\%$ & 3670 $\pm$ 503 \\ 
\hline 
SM $t\bar{t}$ & 16911 $\pm$ 2163 \\ 
\hline 
{\it W}+jets & 242 $\pm$ 52 \\ 
\hline 
{\it Z}+jets & 29 $\pm$ 5 \\ 
\hline 
Single top & 463 $\pm$ 50 \\ 
\hline 
Dibosons & 5 $\pm$ 1 \\ 
\hline 
Total background & 17651 $\pm$ 2164 \\ 
\hline 
\hline 
Data  & 17759 \\ 
\hline 
\hline 
\end{tabular}

%% file: htocsbar.bbl
\begin{thebibliography}{99}
\bibitem{Aad:2012tfa}
G. Aad {\it et al.} (ATLAS Collaboration), Phys.\ Lett.\ B {\bf 716}, 1 (2012).

\bibitem{Chatrchyan:2012ufa}
S. Chatrchyan {\it et al.} (CMS Collaboration), Phys.\ Lett.\ B {\bf 716}, 30 (2012).

\bibitem{Achard:2003gt}
P. Achard {\it et al.} (L3 Collaboration), Phys.\ Lett.\ B {\bf 575}, 208 (2003). 

\bibitem{Heister:2002ev}
A. Heister {\it et al.} (ALEPH Collaboration), Phys.\ Lett.\ B {\bf 543}, 1 (2002).

\bibitem{PhysRevD.80.015017}
M. Aoki, S. Kanemura, K. Tsumura, and K. Yagyu, Phys.\ Rev.\ D {\bf 84}, 055028 (2011).

\bibitem{atlashtocs}
G. Aad, {\it et al.} (ATLAS Collaboration), Phys.\ J.\ C {\bf 736}, (2013).

\bibitem{CMS-HIG-13-035}
CMS Collaboration, CMS-PAS-HIG-13-035, \emph{http://cds.cern.ch/record/1728343}

\bibitem{detectorCMS}
S. Chatrchyan {\it et al.} (CMS Collaboration), JINST {\bf 3}, S08004 (2008).

\bibitem{Alwall:2011uj}
J. Alwall {\it et al.}, JHEP {\bf 1106} (2011) 128. 

\bibitem{Sjostrand:2006za}
T. Sjostrand, S. Mrenna and P.~Z. Skands, JHEP {\bf 0605} (2006) 026.

\bibitem{z2tune}
S. Chatrchyan {\it et al.} (CMS Collaboration), JHEP {\bf 1109} (2011) 109.

\bibitem{cteq}
J. Pumplin {\it et al.}, JHEP {\bf 0207} (2002) 012.

\bibitem{powheg}
S. Alioli, P. Nason, C. Oleari and E. Re, JHEP {\bf 0909} (2009) 111;
E. Re, Eur.\ Phys.\ J.\ C {\bf 71} (2011) 1547.

\bibitem{fewz}
Y. Li and F. Petriello, Phys.\ Rev.\ D {\bf 86} (2012) 094034.

\bibitem{mcfm}
N. Kidonakis, arXiv:1205.3453 [hep-ph].

\bibitem{toppt}
CMS Collaboration, CMS-PAS-TOP-12-027, \emph{http://cds.cern.ch/record/1523611}

\bibitem{Chatrchyan:2011ds}
S. Chatrchyan {\it et al.} (CMS Collaboration), JINST {\bf 6}, P11002 (2011).

\bibitem{CLs}
L. Read, J.\ Phys.\ G: Nucl.\ Part.\ Phys., {\bf 28} (2002) 2693. 

\bibitem{Junk:1999kv}
T. Junk, Nucl.\ Instrum.\ Meth.\ A, {\bf 434} (1999) 435.


\end{thebibliography}
